# The Birth of the HST Snapshot Programs

John Bahcall, Institute for Advanced Study, Princeton

The Snapshot program originated in a lunchtime conversation between Rodger Doxsey and myself in the STScI cafeteria sometime in the spring of 1989. We were both late to lunch and probably were the only people in the cafeteria. The principal topic of conversation was the expected low observing efficiency of the HST. Rodger described the extraordinary difficulty in making a schedule that would use a reasonable percentage of the available time for science observations. Slewing was slow and changing instruments or modes of observing was time consuming. Also, the scheduling software that existed in 1989 was not very powerful.

I asked Rodger, without thinking very carefully about what I was saying, if it would be possible for the software he was developing to insert new objects in the holes in the schedule. I wondered aloud if one could improve the efficiency by choosing new objects, close to the directions of the scheduled targets, from a previously prepared list of interesting objects scattered over the sky. I remember that Rodger suddenly became very quiet, thought about the question, and finally replied something like: "In principle, it is possible." The Snapshot program was born at that lunch.

The subsequent early history was stormy. I proposed the concept of a "Non-Proprietary Survey" at an HST science working group meeting shortly thereafter. All hell broke loose. The instrument PI's and other principal GTOs (guaranteed time observers, I was an Observatory Scientist) all argued heatedly that this project would use up precious HST resources, which were in critically short supply before launch. The other members of the working group felt that STScI resources should be focused on the high priority GTO and GO (general observer) programs, not on some subsidiary (albeit non-proprietary) program.

Riccardo Giacconi, the first STScI director, saved the program from infanticide. Calming the waters, Riccardo persuaded everyone to allow him to work with me and with Rodger to see if we could develop a program in which all the work would be done in Princeton and no STScI resources would be required.

With the enthusiastic help of Rodger, Jim Gunn, Opher Lahav, and Don Schneider, I drew up a science program (HST Program 2775, which subsequently had many aliases including 3034, 3092, …) which proposed WFPC images of relatively bright quasars (463 objects), large angular diameter, peculiar, and interacting galaxies from the ESO and UGC catalogs (402 objects), and selected star fields (17 areas). We had several meetings in the summer and winter of 1989 to set the ground rules for the project. Peter Stockman summarized the results of these meetings in a memo dated August 7, 1989.

The conditions for the implementation of the program may seem stringent by today's operating standards. We agreed that Don and I would search all fields for bright stars that might affect subsequent observations, that we would provide the software and algorithms for feeding the objects to the scheduling system, that we would process and annotate each data tape (including object description, image quality, and science), and that we would deliver the annotated tapes for public distribution every three months. Rodger agreed to develop the capability to assign generic parallel observations in the scheduling system. The Non-Proprietary program was to be assigned the lowest scheduling priority.

In a January 5, 1990 meeting with Rodger, Duccio Macchetto, Larry Petro, and Peter Stockman, we agreed that all of the exposures would be made on gyro control, with no guide stars. This decision was motivated by our desire to have only the least impact on the overloaded STScI resources.

The "gyro-only" policy had a far-reaching science implication that we did not anticipate at the time. We removed the star fields, since they required longer exposures. The science team decided to replace the star fields by exposures of bright but distant ($Z > 1$) quasars. Originally only a small part of the Non-Proprietary survey, the distant quasars were slated to become our primary science program after the mirror problem was found.

The Non-Proprietary Survey, which was dubbed the Non-Proprietary "Snapshot Survey" by (I think ) Peter Stockman, was approved by Riccardo for Director's discretionary time on a trial basis for the early HST observations. Riccardo felt that he had fulfilled his commitment to the HST working group not to use significant STScI resources for the project. We were awarded a magnanimous grant of $20,000 to prepare the target lists, measure positions and magnitudes of all objects, develop the required

software, measure and report regularly on the gyro performance and the telescope pointing accuracy, do the science analysis, publish our results, and deliver the annotated tapes.

The situation changed drastically when, after launch in April 1990, spherical aberration was discovered in the HST images. The wide-field images of galaxies no longer made sense. But, after discussions among the science team, we realized that we could still do a gravitational lens survey of bright but distant quasars, using the sharp core of the PSF to look for close, multiple images. Riccardo allowed me to revise our Director's discretionary time proposal and the Snapshot lensing survey became one of the principal early programs of HST. Dani Maoz was hired as a postdoc at the Institute for Advanced Study, assuming responsibility for the initial technical and scientific analysis.

Our survey played a minor but useful role in the thrilling, frustrating, and stressful early days of bringing the HST Observatory into routine science observations. We obtained frequent observations under standard conditions (same filters, same observing time, similar objects) in the pre-Cycle 0 phase as part of the Science Assessments Tests program. Dani measured large telescope pointing errors (median error 25 arcseconds) and large image drift rates during the exposures. These were traced to the fact that corrections for the effect of stellar aberration had not been activated in the pointing and guiding software for the gyro-only mode. Once this was fixed, it brought about a large reduction in failed target acquisitions in other HST programs. In observing Cycles 0, and 1, we obtained many valuable science observations as the Observatory performance improved. Snapshot observations were scheduled almost routinely.

The Snapshot survey for gravitational lenses was initially described in Bahcall et al., *ApJ* , **387**, 56-68 (1992) and summarized, following a series of other papers in Maoz et al. *ApJ*, **409**, pp. 28-41 (1993). The survey included a total of 498 quasars (as well as star count data) and provided (in addition to other significant scientific results) the first systematic measurement of the frequency of lensing among a large sample of bright quasars, especially in the subarcsecond image regime that only HST could probe.

In observing Cycle 2, STScI announced the Snapshot survey mode as a standard observing option. By this time, the process of finding guide

stars for targets had become computer intensive
rather than personnel intensive. As a result, Snapshot proposals were permitted to make use of guide stars and could therefore cover a wider range of science programs. Today, Snapshot surveys are frequently used and contribute to HST's effectiveness. I am glad that Rodger and I were late for lunch on that spring afternoon in 1989[1].

---

[1] Rodger Doxsey, Duccio Machetto, Larry Petro, and Peter Stockman, are still providing scientific and technical expertise at STScI. Dani Maoz and Don Schneider have moved on from postdoctoral memberships at IAS to faculty positions at Tel Aviv University and Pennsylvania State University, respectively.